\documentclass[twocolumn,showpacs,preprintnumbers,amsmath,amssymb]{revtex4}

\usepackage{graphicx}% Include figure files
\usepackage{dcolumn}% Align table columns on decimal point
\usepackage{bm}% bold math
\usepackage{epsfig}

\newcommand{\be}{\begin{equation}}
\newcommand{\ee}{\end{equation}}
\newcommand{\beq}{\begin{eqnarray}}
\newcommand{\eeq}{\end{eqnarray}}

\newcommand{\W}{\Omega}

\newcommand{\bnn}{\begin{eqnarray*}}
\newcommand{\enn}{\end{eqnarray*}}

\begin{document}
%
%\preprint{APS/123-QED}
%
\title{Intensity correlation and anticorrelations in coherently
prepared atomic vapor}
\author{
    Gombojav O. Ariunbold$^{1,4,*}$,
    Vladimir A. Sautenkov$^{1,3}$,
    Yuri V. Rostovtsev$^{1}$,
    and
    Marlan O. Scully$^{1,2}$}
\affiliation{$^{1}$Institute for Quantum Studies and Department
of Physics, Texas A $\&$ M University, College Station, Texas
77843, USA
\\
$^{2}$ Department of Aerospace and Mechanical Engineering,
Princeton University, Princeton, New Jersey 08544, USA
\\
$^3$ Lebedev Institute of Physics, Moscow 119991, Russia
\\
$^{4}$ Theoretical Physics Laboratory, National University of
Mongolia, 210646 Ulaanbaatar, Mongolia}
\date{\today}
%%%%%%%%%%%%%%%%%%%%%%%%%%%%%%%%%%%%%%%%%%%%%%%%%%%%%%%%%%%%%%%%%%%%%%%%%
%
\begin{abstract}
Motivated by the recent experiment [V.A. Sautenkov, Yu.V.
Rostovtsev, and M.O. Scully, Phys. Rev. A 72, 065801 (2005)], we
develop a theoretical model in which the field intensity
fluctuations resulted from resonant interaction of a dense atomic
medium with laser field having finite bandwidth. The
intensity-intensity cross correlation between two circular
polarized beams can be controlled by the applied external
magnetic field. A smooth transition from perfect correlations to
anti-correlations (at zero delay time) of the outgoing beams as a
function of the magnetic field strength is observed. It provides
us with the desired information about decoherence rate in, for
example, $^{87}$Rb atomic vapor.
\end{abstract}
 % for quantum optics
\maketitle
%%%%%%%%%%%%%%%%%%%%%%%%%%%%%%%%%%%%%%%%%%%%%%%%%%%%%%%%%%%%%%%%%%%%%%%

\section{Introduction}

The fundamental limits of spectral resolution and sensitivity of spectroscopic
techniques, the information transfer and computation rates, spatial resolution
of optical microscopy and imaging are determined by statistical properties of
light. For the last five decades enormous theoretical and experimental
research activities have been devoted to studying fluctuations in classical
and quantum systems~\cite{szbook}.

The first experiment on the
correlation between the intensity fluctuations recorded at the
two different photo-detectors illuminated by the same thermal
light source was performed by Hanbury-Brown and
Twiss~\cite{hanbury56}. In their experiment, photon bunching, i.e.,
an enhancement in the intensity-intensity correlations has been
observed.

Quantum formulation of optical coherences was introduced
by Glauber in his pioneering work~\cite{glauber63}.
Photon anti-bunching has been predicted by Carmichael et
al.~\cite{carmichael76} and then it was firstly observed in
resonance fluorescence experiment by Kimble et
al.~\cite{kimble77}.

As a generalization of the results obtained for two-level atomic
systems~\cite{mandelwolfbook,kimble77a} to fluorescence from
a $\Lambda$ three-level atomic system showing
an anti-bunching effect in the second order correlations has been studied
in~\cite{agarwal79}. Due to four-wave mixing in cold atoms~\cite{braje04}
under condition of electromagnetically
induced transparency (EIT)~\cite{harris97},
Harris and
co-workers~\cite{balic05} have measured the correlation between
Stokes and anti-Stokes photons emitted from Rb atoms with short
time delay.

Kuzmich et al.~\cite{kimble03} have demonstrated a generation of
pair photons with controllable time delay in the issue of quantum
information storage and
retrieval~\cite{chuangnielsenbook,lukin03} using ensemble of
atoms~\cite{duan01}. The correlated photons have been greatly
attracted in the study of, e.g., entanglement
amplifier~\cite{scully05}, subnatural
spectroscopy~\cite{scully95}, quantum microscopy~\cite{scully04},
nonclassical imaging of trapped ions~\cite{agarwal04} and many
others.

A transition from anti-bunching to bunching of light
emitted from a few atoms in a very high finesse cavity has been
demonstrated~\cite{rempe05}. The transition occurs by
increasing a number of atoms interacting with light~\cite{carmichael78}.

The matched fields treated both
classically~\cite{harris93} and quantum
mechanically~\cite{agarwal93} can be another promising
theoretical approach to the switching of correlations in a
three-level atomic sample. However, the result is very sensitive
to the detuning between driven fields and atomic levels.
Photon bunching in the
intensity-intensity correlations between pump and probe fields
for different probe detunings in Rb vapor~\cite{alzar03} and in
the temporal correlations between forward and backward
anti-Stokes photons scattered from sodium vapor~\cite{motomura05}
has been demonstrated.

The most recently, Scully and co-workers~\cite{volodya} have
obtained a smooth transition from EIT correlated to
anti-correlated photons emitted from coherently prepared
$^{87}$Rb vapor.

In the present work we develop a theory to explain the results of
the previous experiments~\cite{volodya}, whereas laser source to
be considered here with a finite bandwidth. A diode laser used in
our experiment, would have low intensity fluctuations but
non-negligible phase fluctuations under certain condition. The
fluctuations of the input light after interacting with the atomic
sample can be enhanced and contain information about atomic
sample. For instance, this has been used as a spectroscopic
tool~\cite{yabuzaki91,walser94}. Particularly, the laser phase
fluctuations can be converted into the intensity fluctuations due
to the interaction of field with atomic
vapor~\cite{mcintyre93,camparo98,bahoura01,martinelli04}.

Furthermore, based on the numerical results, we suggest a new
promising method to estimate decoherence rate for Zeeman
sub-levels.

This paper is organized as follows. In the next
section, the experimental setup is described and the obtianed
results are reported. In section III, we study
the absorption induced fluctuations
of laser beam intensities and their correlations by
considering a generic three-level $\Lambda$ atomic system interacting with
laser fields with orthogonal polarizations.
We obtain the approximate analytical solutions
elucidating an origin of perfect correlations as well as
anti-correlations between two modes. Then, in the next section,
we solve numerically equations of motion to prove the
results of the analytical predictions. The last section is conclusion.
%%%%%%%%%%%%%%%%%%%%%%%%%%%%%%%%%%%%%%%%%%%%%%%%%%%%%%%%%%%%%%%%%%%%%%%%%%%%%
%
\section{Experimental setup and obtained results}
%
%%%%%%%%%%%%%%%%%%%%%%%%%%%%%%%%%%%%%%%%%%%%%%%%%%%%%%%%%%%%%%%%%%%%%%%%%%%%%
\begin{figure}[!ht]
\begin{center}
\includegraphics[width=60mm]{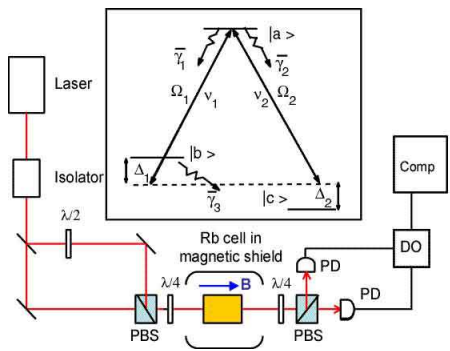} \label{arischeme}
\caption{A simplified schematics of experimental setup and a level
scheme considered for Rb atoms. PBS: polarizing beam-splitter,
$\lambda/2$, $\lambda/4$: wave plates, PD: photo-detector, DO:
digital oscilloscope, Comp: computer.}
\end{center}
\end{figure}
%%%%%%%%%%%%%%%%%%%%%%%%%%%%%%%%%%%%%%%%%%%%%%%%%%%%%%%%%%%%%%%%%%%%%%%%%%%%%

A setup of the experiment (similar to one in ~\cite{volodya}) is
shown in Fig.~1. An external-cavity diode laser~\cite{vassiliev}
is tuned to $D_1$ line ($5S_{1/2} (F=2)\leftrightarrow
5P_{1/2}(F'=1)$) of $^{87}$Rb. An input beam is separated by a
beam-splitter. The polarizations of these two separated beams
become orthogonal using a $\lambda/2$-wave plate put on the way
of one beam and these are combined together by a polarizing
beam-splitter (PBS). After the $\lambda/4$ wave-plate the beam is
a combination of two circular polarized optical fields. A glass
cell of length $L=7.5$cm with Rb vapor (natural abundance) at
density approximately $10^{12}$cm$^{-3}$ is installed in a
two-layer magnetic shield. A simplified level scheme is depicted
in inset of Fig.~1. The opposite circular polarized beams
interact with the vapor and induce a ground state Zeeman
coherence in Rb atoms. EIT resonance is presented in Fig.~2.
%
%%%%%%%%%%%%%%%%%%%%%%%%%%%%%%%%%%%%%%%%%%%%%%%%%%%%%%%%%%%%%%%%%%%%%%%%%%%%%%%%
\begin{figure}[!ht]
\begin{center}
\includegraphics[width=60mm]{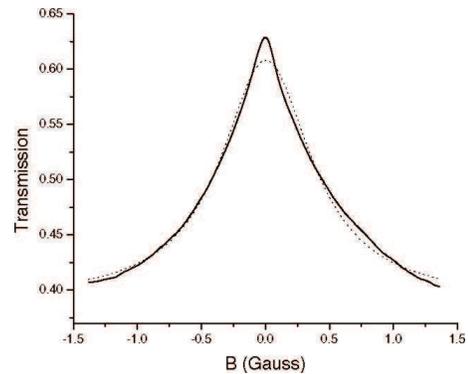} \label{trans}
\caption{A transmission of the optical field through the Rb cell
as a function versus magnetic field $B$ (solid curve). Dots stand
for Lorentzian fit. The total optical power at all at entrance
window is $1mW$.}
\end{center}
\end{figure}
%%%%%%%%%%%%%%%%%%%%%%%%%%%%%%%%%%%%%%%%%%%%%%%%%%%%%%%%%%%%%%%%%%%%%%%%%%%%%%%%
%
Transmitted laser beams after the second $\lambda/4$ wave-plate
are separated again by another polarizing beam-splitter and
focused on fast photodiods (PD) with frequency bandwidth
$75$kHz$-1.2$GHz. The optical path lengths for both beams are the
same. Signals from PDs are sent to a digital oscilloscope (DO). As
varying a magnitude of longitudinal magnetic field the
transmitted fields are changed at the optical power of $0.5$mW
(total power $1$mW) and beam diameter of $0.1$cm for each beam at
the entrance window of the Rb cell.
%
%%%%%%%%%%%%%%%%%%%%%%%%%%%%%%%%%%%%%%%%%%%%%%%%%%%%%%%%%%%%%%%%%%%%%%%%%%%%%%%%
\begin{figure}[!ht]
\begin{center}
\includegraphics[width=60mm]{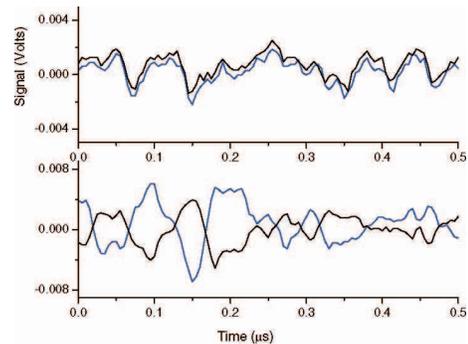} \label{intensityexp}
\caption{Waveforms from photo-detectors with total optical power
of laser beams at front window of Rb cell is $1$mW. The
fluctuations of intensities versus time for two coinciding beams
with (a) no magnetic field $B=0$ and (b) applied magnetic field
$B=-0.47$Gauss.}
\end{center}
\end{figure}
%%%%%%%%%%%%%%%%%%%%%%%%%%%%%%%%%%%%%%%%%%%%%%%%%%%%%%%%%%%%%%%%%%%%%%%%%%%%%%%%
%
The time dependent intensity fluctuations $\delta I_{1,2}(t)$ of
both optical beams transmitted through Rb vapor, are registered by
the photodetectors (see, Fig.~3). Data presented here is a part
of the recorded data in $10\mu$sec. The signal in Volts is
proportional to laser intensity as $500$V/W. Furthermore, the
intensity-intensity correlations between two modes can be
calculated using the observed data for the intensity
fluctuations. The second order correlation function
$G^{(2)}(\tau)$ for intensity fluctuations of two optical beams
with time delay $\tau$ is given by
\begin{equation} \label{classicalcorr}
G^{(2)}(\tau)=\frac{\langle \delta I_1(t) \delta
I_2(t+\tau)\rangle}{\sqrt{\langle [\delta I_1(t)]^2 \rangle
\langle [\delta I_2(t+\tau)]^2 \rangle}}
\end{equation}
where the time average of arbitrary variable $Q(t)$ is defined as
$\langle Q(t) \rangle=\int_t^{t+T}Q(t)dt/T$. The integration time
$T$ is taken to be as large as $10\mu$s. In the absence of the
external magnetic field $B=0$ where EIT condition is fulfilled
(two-photon detuning is zero), the induced fluctuations of the
transmitted beams by Rb vapor are almost synchronized (see
Fig.~3(a)). In the case of zero detuning, the intensity-intensity
correlation curve of Fig.~4(a) has a sharp spike clearly showing
bunching. The magnitude of the correlation peak at $\tau=0$ is of
$0.9$ and the average background is near $0.15$. The width of the
correlation peak increased as reduces the optical power. On the
other hand, the most intriguing feature is observed when an
applied magnetic field is of $B=-0.47$Gauss. As is seen from
Fig.~3(b), intensity fluctuations are out of phase when two
photon detuning becomes non zero.
%
%%%%%%%%%%%%%%%%%%%%%%%%%%%%%%%%%%%%%%%%%%%%%%%%%%%%%%%%%%%%%%%%%%%%%%%%%%%%%%%%
\begin{figure}[!ht]
\begin{center}
\includegraphics[width=60mm]{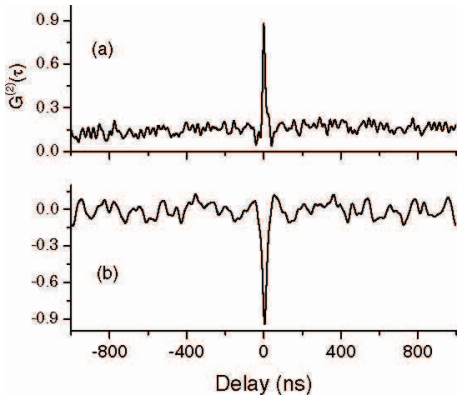} \label{correxp}
\caption{Intensity-intensity correlation functions
$G^{(2)}(\tau)$ as functions of time delay $\tau$ for magnetic
fields $B=0$ (above) and $B=-0.47$ (bottom).}
\end{center}
\end{figure}
%%%%%%%%%%%%%%%%%%%%%%%%%%%%%%%%%%%%%%%%%%%%%%%%%%%%%%%%%%%%%%%%%%%%%%%%%%%%%%%%
%
A presence of the magnetic field demonstrates an exhibition of
anticorrelation (see Fig.~4(b)) as is expected from the data
plotted in Fig.~3(b). The width of the peaks is associated with
the saturated width of resonance in Rb vapor absorption (a single
photon resonance)~\cite{akulshin90}. Moreover, the second order
correlation $G^{(2)}(\tau)$ obtained with spatially separated
beams (distance between beams $0.3$cm which is bigger than the
beam diameter $0.1$cm) has a correlation peak (at $\tau=0$) of
$0.7$ and a larger background of $0.3$. This contrast indicates
that the enhanced correlations are clearly due to overlap of two
beams. We have also performed a set of measurements of
$G^{(2)}(\tau=0)$ for different values of the magnetic field $B$
at optical power $1$mW (see, Fig.~10). A measurement shows that a
perfect switching from photon correlation to anticorrelation
where the correlation peak with magnitude $0.9$ for zero magnetic
field $B=0$ switches to a dip with magnitude $-0.9$ for
$B=-0.47$Gauss. This pronounced modification of waveforms is very
important in determination of some experimental parameters which
will be shown later. The width related to intensity-intensity
correlations $G^{(2)}(\tau=0)$ are $0.24$Gauss, almost four times
narrower than the corresponding EIT width which is $0.85$Gauss
for optical power $1$mW.
%%%%%%%%%%%%%%%%%%%%%%%%%%%%%%%%%%%%%%%%%%%%%%%%%%%%%%%%%%%%%%%%%%%%%
%
\section{Theoretical results}

\subsection{Three-level $\Lambda$ atoms driven by laser with finite
bandwidth}

Let us consider interaction of two modes generated by the
diode laser with three-level $\Lambda$ atoms that have a level scheme
shown in Fig.~1. The equations of motion for this system are
given by (e.g. see~\cite{agarwal79})
\begin{eqnarray} \label{main}
\dot{{\rho}}_{bc} & = & -(\gamma_3+i \Delta)
{\rho}_{bc}+i\Omega_1 {\rho}_{ca}^\dagger- i\Omega_2 \rho_{ba}
\nonumber\\
\dot{{\rho}}_{ba} & = & -(\gamma_1+i \Delta_1){\rho}_{ba}-i
\Omega_1({\rho}_{bb}-{\rho}_{aa})
\nonumber\\
&&-i\Omega_2{\rho}_{bc}+i\dot{\phi}(t)\rho_{ba}
\nonumber\\
\dot{{\rho}}_{ca} & = & -(\gamma_2-i \Delta_1) {\rho}_{ca}-i
\Omega_2({\rho}_{cc}-{\rho}_{aa})
\nonumber \\
&& -i\Omega_1 {\rho}_{bc}^\dagger+i\dot{\phi}(t)\rho_{ca}
\nonumber\\
\dot{{\rho}}_{bb} & = & \tilde{\gamma}_3 {\rho}_{cc}
+\tilde{\gamma}_1{\rho}_{aa}+i\Omega_1
{\rho}_{ba}^\dagger-i\Omega_1 {\rho}_{ba}
\\
\dot{{\rho}}_{cc} & = & \tilde{\gamma}_2{\rho}_{aa}
-\tilde{\gamma}_3{\rho}_{cc}-i\Omega_2{\rho}_{ca}+i\Omega_2{\rho}_{ca}^\dagger\nonumber
\end{eqnarray}
where ${\rho}_{aa}=1-{\rho}_{bb}-{\rho}_{cc}$. The single and two
photon detunings are
$\Delta_{1}=\omega_a-\omega_c-\nu_2=\nu_1-\omega_a+\omega_b$,
($\nu_1=\nu_2$) and $\Delta=\omega_b-\omega_c$. The effective
decay parameters defined~\cite{roos03} as
$\gamma_3=(\tilde{\gamma}_3+\tilde\gamma_{21}+\tilde\gamma_{12})/2$,
$\gamma_1=(\tilde{\gamma}_1+\tilde{\gamma}_2+\tilde\gamma_{21}+\tilde\gamma_{31})/2$
and
$\gamma_2=(\tilde{\gamma}_1+\tilde{\gamma}_2+\tilde{\gamma}_3+\tilde\gamma_{12}+\tilde\gamma_{32})/2$,
where $\tilde\gamma_{1}$ and $\tilde\gamma_2$ correspond to the
spontaneous emission rates from level $|a\rangle$ to levels
$|c\rangle$ and $|b\rangle$, respectively; $\tilde\gamma_{21}$,
$\tilde\gamma_{12}$, $\tilde\gamma_{32}$ and $\tilde\gamma_{31}$
stand for dephasing rates, and $\tilde\gamma_3$ is population
decay rate of the level $|b\rangle$. In derivations of
Eq.(\ref{main}), we have kept the operator normal ordering, i.e.,
we neglect rapidly oscillating terms~\cite{mandelwolfbook} and
used re-scaled variables as ${\rho}_{ij} \rightarrow 1/N_c
{{\rho}}_{ij}$, $N_c$ is number of collective atoms. A diode
laser radiation experiences phase diffusion, and the phase
$\phi(t)$ in Eq.(\ref{main}) represents the fluctuating
phase~\cite{kimble77a} of driven field which is characterized by
Wiener-Levy diffusion process~\cite{gardnerzollerbook}. For such
process average and two-time correlation function of stochastic
variables are given by
\begin{eqnarray} \label{white}
\langle\overline{\dot\phi(t)}\rangle & = & 0 \nonumber\\
\langle\overline{\dot\phi(t)\dot\phi(t')}\rangle & = &
2D\delta(t-t')
\end{eqnarray}
where $D$ is the diffusion coefficient; the stochastic averages
denoted by the upper bar. Thus, the input laser field has a
Lorentzian spectrum with a FWHM bandwidth of $D/\pi$Hz. In a
realistic situation, the phase correlation has a finite
relaxation time. The Gaussian process in which the correlations
are determined by the exponential function of time delay is often
referred to as Ornstein-Uhlenbeck~\cite{ornstein54} or colored
noise. A stationary equation after taking stochastic average of
Eq.(\ref{main}) is shown in Appendix. The numerical simulations
of Eq.(\ref{main}) will be presented below.

\subsection{Absorption induced intensity-intensity correlations}

Propagation equations for the laser fields are given by \be
{\partial\W_1\over\partial z}=i\kappa_1\rho_{ab}, \;\;\;
{\partial\W_2\over\partial z}=i\kappa_2\rho_{ac}. \ee

In order to give a qualitative theoretical analysis of our
experimental results, let us adopt a theory which implies for a
thin absorbing medium. It is assumed that the transmitted field
could be understood as a superposition of input and induced
fields in the first order approximation for $\kappa_{1,2} L$, if
$\kappa_{1,2}L \ll 1$; here $L$ is the length of the atomic
sample and $\kappa_{1,2}$ are some coefficients~\cite{walser94}.
Furthermore, this could be still valid for a preferably long
medium with a weak absorption, but satisfying the condition
$\kappa_{1,2} L\ll 1$. In what follows, we will show that this
approximation reproduces the observed results qualitatively.
Following Walser et al.~\cite{walser94} and Martinelli et
al.~\cite{martinelli04}, the transmitted fields are given by
\begin{eqnarray} \label{output}
\Omega_{1}^{out}(t) &\approx& \Omega_{1} + i(\kappa_1 L)
\rho_{ac}(t), \nonumber\\
\Omega_{2}^{out}(t) &\approx& \Omega_{2} + i(\kappa_2 L)
\rho_{ab}(t).
\end{eqnarray}
To include Doppler effect, the coherence terms in
Eq.(\ref{output}) should be averaged by the Maxwell-Boltzmann
velocity distribution. However, the Doppler broadening may play
important role in many other experiment with Rb atomic vapor,
but, in this case, it turns out not to be so crucial,
because we are interested only in correlation behaviour as
functions of two-photon detuning, instead of one-photon detuning.
The transmitted intensities $I_{1,2}(t) \Rightarrow|
\Omega_{1,2}^{out}(t)|^2$ are, thus, written as
\begin{eqnarray}
I_1(t) &\approx& \Omega_{1}^2 + \Omega_1 (\kappa_1 L)
{\rm Im}\{\rho_{ac}(t)\}\nonumber\\
I_2(t) &\approx& \Omega_{2}^2 +\Omega_2 (\kappa_2 L) {\rm Im}
\{\rho_{ab}(t)\}.
\end{eqnarray}
here $\rho_{aq}$ is the atomic coherence term for level $a$ and
$q$, ($q=b,c$) and we assume that the input fields are real and
much stronger than the induced ones. Defining that $\delta
Q(t)=Q(t)-\overline{Q(t)}$ stands for the fluctuation of
arbitrary variable $Q(t)$, the intensity fluctuation to be read
\begin{eqnarray} \label{intensityfluctuation}
\delta I_1(t) & = & \Omega_1 (\kappa_1 L) {\rm Im}\{\delta
\rho_{ac}(t)\}\nonumber\\
\delta I_2(t) & = & \Omega_2(\kappa_2 L) {\rm Im}\{\delta
\rho_{ab}(t)\}.
\end{eqnarray}
From this expression, it is easy to check that $\langle \delta
I_{1,2}(t)\rangle=0$, because $\langle \delta
\rho_{aq}(t)\rangle=0$, $q=c,b$. From
Eq.(\ref{intensityfluctuation}), using the definition
Eq.(\ref{classicalcorr}) one obtains the intensity-intensity
correlations
\begin{equation} \label{corr1}
G^{(2)}(\tau)=\frac{\langle {\rm Im}\{\delta \rho_{ac}(t)\} {\rm
Im}\{\delta \rho_{ab}(t+\tau)\}\rangle}{\sqrt{\langle [{\rm
Im}\{\delta \rho_{ac}(t)\}]^2 \rangle \langle [{\rm Im}\{\delta
\rho_{ab}(t)\}]^2 \rangle}}
\end{equation}
The undertaking process is stationary, thus, the argument $t+\tau$
of the last term in the denominator is displaced by $t$.

%%%%%%%%%%%%%%%%%%%%%%%%%%%%%%%%%%%%%%%%%%%%%%%%%%%%%%%%%%%%%%%%%%%%%%%%%%%%%%%%%%%%%%%
%
%
%
\subsection{Approximate theoretical analysis}
\subsubsection*{Zero detuning}
Let us consider the resonant case where all detunings are set to
be zero and $\Omega_1=\Omega_2$, $\tilde{\gamma}_3=0$,
$\gamma_1=\gamma_2$. It is easy to show analytically that
$\langle\overline\rho_{bc}\rangle=\langle\overline\rho_{cb}\rangle$
where an equation for stationary state $\overline\rho_{bc}$ is
given in Appendix. According to exact numerical simulations of
Eq.(\ref{main}), the coherence term $\rho_{bc}$ are real i.e.,
$\rho_{bc}\cong \rho_{bc}^\dagger$. Therefore, equations for
coherence terms $\rho_{ba}$ and $\rho_{ca}$ can have symmetrical
forms as
\begin{eqnarray} \label{decoupled}
\dot{{\rho}}_{ba} & = & -\gamma_1{\rho}_{ba}-i\Omega_1
({\rho}_{bb}-\rho_{aa})
+i\dot{\phi}(t)\rho_{ba}-i\Omega_1\rho_{bc}
\nonumber\\
\dot{{\rho}}_{ca} & = & -\gamma_1 {\rho}_{ca}-i\Omega_1
({\rho}_{cc}-\rho_{aa})+i\dot{\phi}(t)\rho_{ca}-i\Omega_1\rho_{bc}
\nonumber\\
\end{eqnarray}
From Eq.(\ref{decoupled}), one can see that two equations, thus,
two modes are decoupled. Note that because of symmetrical atomic
configuration, it is obvious that populations $\rho_{bb}$ and
$\rho_{cc}$ are identical $\rho_{bb}\approx \rho_{cc}$. This is,
of course, true only in a resonant case. From Eq.(\ref{decoupled}),
it follows that
\begin{equation}
\rho_{ba}(t)\thickapprox\rho_{ca}(t)
\end{equation}
This correlated behavior could be understood as follows. Phase
fluctuations of incident beams are converted into intensities
fluctuations via atom-field interactions as is seen from
Eq.(\ref{main}). Roughly speaking, the three-level atoms would
experience driven fields with the effective Rabi frequencies
fluctuating around $\Omega_1$ which is resonant to the degenerate
lower levels. Because the noise contribution is the same in two
modes, any instant deviations from the resonance condition will be
also the same. Thus, the induced absorption should be also the
same for both modes. In this sense, this system would be in close
relation to what is called a correlated emission laser firstly
proposed by Scully~\cite{szbook}, in which pairs of induced
photons of different modes can be generated simultaneously
exhibiting a sharp bunching.
%%%%%%%%%%%%%%%%%%%%%%%%%%%%%%%%%%%%%%%%%%%%%%%%%%%%%%%%%%%%%%%%%%%%%%%%%%%%%%%%
%
%
%
\subsubsection*{Non-zero detuning}
In the non-degenerate situation, the equations for two modes are
coupled. However, the effective Rabi frequencies would have again
the same fluctuations, but, $\Omega_1$ is no longer resonant to
the lower levels. The induced absorption should not be the
equivalent in this case, because, the deviations of the effective
Rabi frequencies would become farther from one of ground levels,
but, closer to another at any instant time. As a consequence, the
populations of the excited and ground states would also
fluctuate. By virtue of $\dot{\rho}_{bb} + \dot{\rho}_{cc}
\approx 0$, since $\dot{\rho}_{aa}\approx 0$, it is possible to
do the following assumption as
\begin{eqnarray}
\rho_{bb}(t) & = & c + f(t),\nonumber\\
\rho_{cc}(t) & = & c - f(t).
\end{eqnarray}
where a fluctuation $f(t)$ is real and $c$ is constant. Note that
it is not necessary to know an explicit expression for $f(t)$.
This coupling function $f(t)$ appears only because of non-zero
detuning, otherwise it is zero. The equations for two
polarizations, $\rho_{ba}$ and $\rho_{ca}$, are given by
\begin{eqnarray}
\dot{{\rho}}_{ba} & = & -(\gamma_1+i
\Delta_1){\rho}_{ba}-i\Omega_1 f(t)
+i\dot{\phi}(t)\rho_{ba}-i\Omega_1\rho_{bc}+c'
\nonumber\\
\dot{{\rho}}_{ca} & = & -(\gamma_1-i\Delta_1) {\rho}_{ca} +i
\Omega_1
f(t)+i\dot{\phi}(t)\rho_{ca}-i\Omega_1\rho_{bc}^\dagger+c'
\nonumber\\
\end{eqnarray}
The formal solutions to be read
\begin{eqnarray} \label{solutions}
{\rho}_{ba}(t) & \sim & +\Phi(t)+{\rm c_1}
\nonumber\\
{\rho}_{ca}(t) & \sim & -\Phi(t)+{\rm c_2}
\end{eqnarray}
where $\Phi(t)=-i\Omega_1\int_{t_0}^t dt'{\rm e}^{-\gamma_1
(t-t')+i(\phi(t)-\phi(t'))}f(t')$, $c_{1,2}=-i\Omega_1\int_{t_0}^t
dt'{\rm e}^{-\gamma_1 (t-t')+i(\phi(t)-\phi(t'))}(\rho_{bc,cb}+c)$
which can be slowly varying for $t_0\Rightarrow -\infty$,
$t\Rightarrow\infty$. Note that, contributions of $\Delta_1$ to
solutions are neglected since, later on, only imaginary part of
amplitudes will be of interest. Eq.(\ref{solutions}) clearly
indicates an exhibition of anti-correlation between two modes.
%%%%%%%%%%%%%%%%%%%%%%%%%%%%%%%%%%%%%%%%%%%%%%%%%%%%%%%%%%%%%%%%%%%%%%%%
%
%
%
\subsection{Numerical results}
In the Ornstein-Uhlenbeck process~\cite{ornstein54}, the colored
noise $\xi(t)$ yields the steady-state correlation function
\begin{equation} \label{colorednoise}
\langle \overline{\xi(t)\xi(t')} \rangle=\Theta\lambda_L {\rm
e}^{-\lambda_L|t-t'|}
\end{equation}
with $\langle \xi(t) \rangle=0$. The stochastic differential
equation Eq.(\ref{main}) can be solved using Monte-Carlo
numerical simulations. A Box-Mueller algorithm and the
Euler-Maruyama method are used to realize the colored noise.
Namely, one can see that the generated noise by the fast, integral
algorithm developed in~\cite{fox88}, is in a perfect agreement
with the analytical definition given by Eq.(\ref{colorednoise}) due to
averaging over as many as 1000 different realizations. A
relaxation time $1/\lambda_L$ is taken to be small, to ensure that
undertaking process would be approximately white noise, i.e.,
$\xi(t)\sim \dot{\phi}(t)$. As a matter of fact, various choices
of parameters $\Theta$ and $\lambda_L$, should not drastically
influence to the final results. The numerical solutions of
Eq.(\ref{main}) allows one to obtain the intensity fluctuations
defined by Eq.(\ref{intensityfluctuation}).
%
%%%%%%%%%%%%%%%%%%%%%%%%%%%%%%%%%%%%%%%%%%%%%%%%%%%%%%%%%%%%%%%%%%%%%%%%%%%%%%%%
\begin{figure}[!ht]
\begin{center}
\includegraphics[width=60mm]{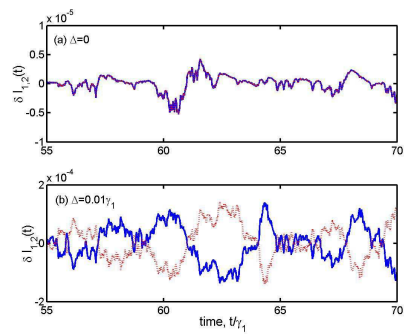} \label{intensity}
\caption{Synchronized dynamics for intensity differences of two
fields (dotted and solid curves) given by
Eq.(\ref{intensityfluctuation}) are depicted for resonant (above
figure, $\Delta=0$) and non-resonant (bottom figure,
$\Delta=0.01\tilde\gamma_1$) cases. All rates and Rabi
frequencies are taken to be
$\tilde\gamma_{12,21}=0.01\tilde\gamma_1$,
$\tilde\gamma_{13,31}=\tilde\gamma_3=0$,
$\tilde\gamma_1=\tilde\gamma_2$ and
$\Omega_1=\Omega_2=\tilde\gamma_1$}
\end{center}
\end{figure}
%%%%%%%%%%%%%%%%%%%%%%%%%%%%%%%%%%%%%%%%%%%%%%%%%%%%%%%%%%%%%%%%%%%%%%%%%%%%%%%%
%
%
The numerical results of Eq.(\ref{intensityfluctuation}) are
plotted by dotted and solid curves in Fig.~5. In resonant case,
dynamics of two modes are in phase i.e., well synchronized. The
dephasing rate for both cases are chosen to be much smaller than
decay rates. Absolute values of two Rabi frequencies are the same
as is considered in the experiment. If two-photon detuning
$\Delta$ becomes comparable to $\gamma_3$ then dynamical
behaviors are absolutely out of phase.
%
%%%%%%%%%%%%%%%%%%%%%%%%%%%%%%%%%%%%%%%%%%%%%%%%%%%%%%%%%%%%%%%%%%%%%%%%%%%%%%%%
\begin{figure}[!ht]
\begin{center}
\includegraphics[width=60mm]{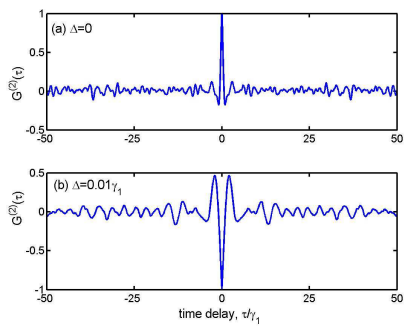} \label{corr}
\caption{A perfect photon correlation (a spike at $\tau=0$, see,
(a)) for $\Delta=0$ and an anti-correlation of two modes (a dip
at $\tau=0$, see, (b)) between two modes for
$\Delta=0.01\tilde\gamma_1$ are obtained. The plots are
calculated for the same parameters as in Fig.~5.}
\end{center}
\end{figure}
It is seen more clearly from Fig.~6. Dynamical features shown in
Fig.~5, can be seen more clearly in terms of cross correlation
functions. Figure~6 describes a switching between two completely
different behaviors, namely, correlation and anti-correlation of
two modes. %The widths of the spike and dip in correlations are
%quite narrow as in Rb vapor~\cite{volodya}.
%%%%%%%%%%%%%%%%%%%%%%%%%%%%%%%%%%%%%%%%%%%%%%%%%%%%%%%%%%%%%%%%%%%%%%%%%%%%%%%%%%%%%%%%%%%%
%
%
%
%
\section{A Decoherence rate determined by switching in intensity-intensity correlations}
In what follows, we analyze this switching in more detail.
Similarly as in~\cite{volodya}, we focus on correlation functions
with zero time delay $G^{(2)}(\tau=0)$. First of all,
Eq.(\ref{corr1}) is obtained for fixed $\Omega_1$ but, different
$\gamma_3$ and shown in Fig.~7. Note that transitions from
correlated photons to anti-correlated ones are appeared to be
smooth and have certain widths.
%
%%%%%%%%%%%%%%%%%%%%%%%%%%%%%%%%%%%%%%%%%%%%%%%%%%%%%%%%%%%%%%%%%%%%%%%%%%%%%%%%
\begin{figure}[!ht]
\begin{center}
\includegraphics[width=60mm]{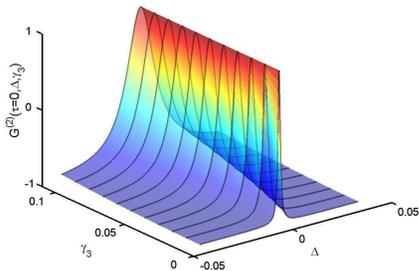} \label{gamma3D}
\caption{3D plot for correlation $G^{(2)}(\tau=0,\Delta,\gamma_3)$
with zero time delay ($\tau=0$) as a function of detuning
$\Delta$ and decoherence rate $\gamma_3$. Rabi frequencies are
$\Omega_1=\Omega_2=\tilde\gamma_1$.}
\end{center}
\end{figure}
These widths are getting more wide with the increase of dephasing
rates.
%
%%%%%%%%%%%%%%%%%%%%%%%%%%%%%%%%%%%%%%%%%%%%%%%%%%%%%%%%%%%%%%%%%%%%%%%%%%%%%%%%
\begin{figure}[!ht]
\begin{center}
\includegraphics[width=60mm]{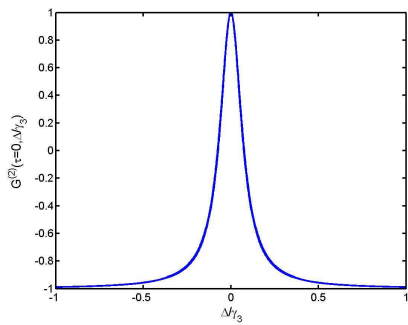} \label{gamma2D}
\caption{A sample of correlation functions
$G^{(2)}(\tau=0,\Delta/\gamma_3)$ with scaled variable as
$\Delta/\gamma_3$ for different decoherence rates:
$\gamma_3=0.01, 0.02, ..., 0.11\tilde\gamma_1$. All curves
coincide. The plots are calculated for the parameters taken from
Fig.~6.}
\end{center}
\end{figure}
Surprisingly, as shown in Fig.~8, the widths are 'invariant' as
functions of re-scaled detuning variable $\Delta/\gamma_3$ with
respect to the corresponding decoherence rates $\gamma_3$.
Moreover, let us test also how a Rabi frequency's change might
affect to correlations.
%
%%%%%%%%%%%%%%%%%%%%%%%%%%%%%%%%%%%%%%%%%%%%%%%%%%%%%%%%%%%%%%%%%%%%%%%%%%%%%%%%
\begin{figure}[!ht]
\begin{center}
\includegraphics[width=60mm]{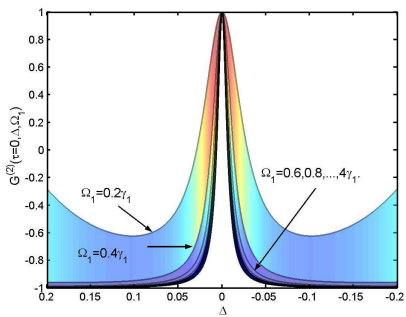} \label{omega}
\caption{2D view of 3D plot for correlation
$G^{(2)}(\tau=0,\Delta,\Omega_1)$ depending on different values of
both detuning and Rabi frequency for fixed dephasing rate
$\gamma_3=0.1\tilde\gamma_1$.}
\end{center}
\end{figure}
In Fig.~9, we depict numerical results for Eq.(\ref{corr1})
depending on not only detuning, but also, Rabi frequencies for
two different fixed values of $\gamma_3$. As a matter of fact,
the correlation curves are again 'invariant' for all Rabi
frequencies those being not smaller $\Omega_1\geq \tilde\gamma_1$.
%
%%%%%%%%%%%%%%%%%%%%%%%%%%%%%%%%%%%%%%%%%%%%%%%%%%%%%%%%%%%%%%%%%%%%%%%%%%%%%%%%
\begin{figure}[!ht]
\begin{center}
\includegraphics[width=60mm]{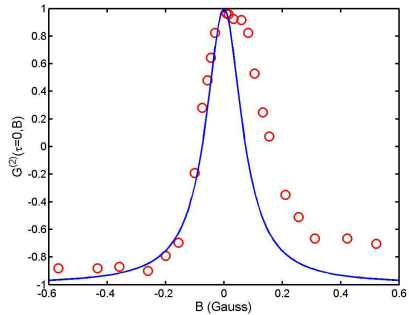} %\label{gamma3}
\caption{A finding dephasing rate in Rb vapor. Correlation
$G^{(2)}(\tau=0,B)$ varying with magnetic field strength $B$ is
compared to the experimental data which allows us to get scaling
factor being $\alpha\sim 1[G]$. Dephasing rate is, thus, found to
be $\gamma_3\sim 1$MHz independently from $\tilde\gamma_1$. No
free parameters are used here. Experimental data is presented by
circles for optical power $1mW$}
\end{center}
\end{figure}
These two intriguing results for the correlation functional
invariance do promise a relatively precise determination of the
decoherence rate in the robust way from the experimental point of
view. As is mentioned in the introduction, the atomic energy
levels are perturbed, due to the interaction of the magnetic
moments of atoms with the external magnetic field $B$. This leads
to a non-degeneracy of atomic ground levels. This shift is given
by
\begin{equation} \label{delta1}
\Delta=a B
\end{equation}
The  constant is defined by Bohr's magneton, the magnetic quantum
numbers and gyromagnetic factor as $a=(\mu_B/\hbar)g(m_2-m_1)$. To
be more explicit, let us concentrate on $^{87}$Rb vapor in
connection to our experiment~\cite{volodya}. An external-cavity
diode laser is tuned to $5S_{1/2} (F=2)\leftrightarrow
5P_{1/2}(F'=1)$. Applied magnetic field leads approximately to
three level $\Lambda$ atomic configuration with a common upper
level $5P_{1/2} (F'=1,m_3=0)$ and two lower levels $5S_{1/2} (F=2,
m_1=-1)$ and $5S_{1/2} (F=2, m_2=+1)$. Using the facts that
$g=0.5$ for $S_{1/2} (F=2)$ and $\mu_B/\hbar=1.4$MHz, the
constant would be estimated $a=1.4$MHz/G. Moreover, from
Eq.(\ref{delta1}), the relation for $B$ can be rewritten as
\begin{equation} \label{B}
B=\alpha\frac{\Delta}{\gamma_3}
\end{equation}
where $\alpha$ is a scaling factor. For fixed $\alpha$, all
correlations with no time delay are supposed to be invariant. As
is seen from Fig.~10, experimental data (circles) are
form preserved and have the identical widths regardless of
optical powers choices. Asymmetry with respect to the zero
detuning may be due to Stark shift which is not of interest in
the present situation. Because of two conditions of invariance,
there is a very good reason to believe that a form of theoretical
correlation functions for all $\Omega_1 \gg \tilde\gamma_1$ and
any of $\gamma_3$, should be equivalent to the experimental data
measured for particular $\Omega_{exp}$ and $\gamma_{3exp}$. In
this spirit, theoretical curves by changing scale $\alpha$, can
be compared with the measurements. Remember that
$\alpha=\gamma_{3exp}/a$, a new formula for dephasing rate can be
given as
\begin{equation} \label{gamma3}
\gamma_{3exp}=\alpha a
\end{equation}
Once $\alpha$ could be found from experimental data, the dephasing
rate for that atomic vapor, is determined by formula
Eq.(\ref{gamma3}). As we expected this is also independent from
population decay rates. For Rb vapor, we have found from Fig.~10,
that scaling factor $\alpha\sim 1[G]$. Therefore, the decoherence
rate is estimated to be $\gamma_{3exp} \sim 1$~MHz.
\section{Conclusion}
An experimental demonstration of intensity correlations and
anti-correlations of coupled fields in a dense Rb vapor is
reported. A lower level coherence is created between Zeeman
sub-levels by two laser beams with orthogonal circular
polarizations. Intensity fluctuations induced by resonant medium
are correlated under resonance EIT condition and anti-correlated
in presence of non-zero two photon detuning. A narrow correlation
peak and anti-correlation dip, in time domain, are associated with
frequencies above EIT width and natural optical width. A
dependence of correlations on magnetic field (two-photon
detuning) show resonance behavior. The resonances are near 4
times narrower than the width of the observed EIT resonances. A
smooth transition from perfect correlations to anti-correlations
(at zero delay time) between the outgoing beams as functions of
the magnetic field strength is robust with respect to a variety
of different choices of physical parameters involved and, thus,
can provide us with the desired information about decoherence in
three level atomic vapor. Moreover, correlation properties of
coupled fields in $\Lambda$ scheme can be used to reduce noise
and improve performance of EIT based atomic clocks and
magnetometers. The phase noise to intensity noise conversion is
an important physical process limiting the accuracy. In EIT
atomic clock and
magnetometers~\cite{fleishhauer94,kitching02,budker02}, it is
possible to avoid the contribution of the atomic medium induced
excess intensity noise.

\section*{Acknowledgements}
The authors thank to S.E. Harris, R. Glauber,  L.V. Keldysh, A.
Muthukrishnan, A. Patnaik, Z.E. Sariyanni, A.V. Sokolov, A.S.
Zibrov, I. Novikova, L. Davidovich for useful and fruitful
discussions, V.V. Vasiliev for his help with external cavity
laser, H. Chen for his help in experiment and gratefully
acknowledge the support from the Office of Naval Research under
Award No. N00014-03-1-0385, the Air Force Research Laboratory
(Rome, NY), Defense Advanced Research Projects Agency-QuIST,
Texas A$\&$M University Telecommunication and Information Task
Force (TITF) Initiative, and the Robert A.\ Welch Foundation
(Grant No. A-1261).
\section*{Appendix: Stochastic averaging of equations with multiplicative noise}
A stochastic average of an arbitrary dynamical variable $F(x)$,
i.e., a path-integral over all possible realizations of the random
numbers $x(t)$, is given by
\begin{equation}
\overline{F(x)}= \int D\mu[x] F(x) \nonumber
\end{equation}
A white noise does satisfy the relations $\overline{x(t)}=0$ and
$\overline{x(t)x(s)} =\Gamma \delta(t-s)$. A functional measure
$D\mu[x]$ has a Gaussian density
\begin{equation}
D\mu[x]=N^{-1}Dx e^{-\frac{1}{2\Gamma}\int d\tau x(\tau)^2}
\nonumber
\end{equation}
here $N$ is normalization coefficient to assure
\begin{equation}
\int D\mu[x]=1 \nonumber
\end{equation}
A characteristic function is found to be
\begin{equation} \label{function}
Z(g)=\overline{e^{i\int d\tau x(\tau) g(\tau)}}
=e^{-\frac{\Gamma}{2}\int d\tau g(\tau)^2}
\end{equation}
Following W$\rm \acute{o}$dkiewicz~\cite{wodkiewicz79}, let us
consider the following stochastic equations
\begin{equation}
\frac{d \Psi}{d t}=M_0 \Psi +i x(t)M\Psi \nonumber
\end{equation}
In the interaction picture where $\Psi_I(t)=e^{-M_0 t}\Psi(t)$,
the equation can be rewritten as
\begin{equation} \label{interaction}
\frac{d \Psi_I}{d t}=i x(t)M_I(t)\Psi_I
\end{equation}
where $M_I(t)=e^{-M_0 t}M e^{M_0 t}$. A formal solution of
Eq.(\ref{interaction}) is given by
\begin{equation}
\Psi_I(t)=T e^{i \int_0^t d\tau x(\tau)
M_I(\tau)}\Psi_I(0)\nonumber
\end{equation}
here T is the time ordering operator. Using the relation
Eq.(\ref{function}), the stochastic average to be read
\begin{eqnarray}
\overline{\Psi}_I(t)&=&\overline{T e^{i \int_0^t d\tau x(\tau)
M_I(\tau)}} \Psi_I(0)\nonumber\\
&=&T e^{-\frac{\Gamma}{2} \int_0^t d\tau M_I(\tau)^2} \Psi_I(0)
\nonumber
\end{eqnarray}
This is equivalent to the equation
\begin{equation}
\frac{d \overline{\Psi}_I }{d t}=-\frac{\Gamma}{2}
M_I(t)^2\overline{\Psi}_I
\end{equation}
and, finally, we arrive at
\begin{equation}
\frac{d \overline{\Psi}}{d t}=M_0 \overline{\Psi}
-\frac{\Gamma}{2} M^2\overline{\Psi}.
\end{equation}
This is the expected stationary equation.

$^*$ e-mail: goa@physics.tamu.edu


\begin{thebibliography}{99}
%


\bibitem{szbook} M.O. Scully and S. Zubairy, {\it Quantum Optics}
(Cambridge University Press, Cambridge, 1997).
%
\bibitem{hanbury56} R. Hanbury-Brown and R.Q. Twiss, {\it Nature}
{\bf 177} 27 (1956).
%
\bibitem{glauber63} R.J. Glauber, {\it Phys. Rev. Lett.} {\bf 130}
2529 (1963).
%
\bibitem{carmichael76} H.J. Carmichael and D.F. Walls, {\it J.
Phys.} B {\bf 9} 1199 (1976).
%
\bibitem{kimble77} H.J. Kimble, M. Dagenais and L. Mandel, {\it Phys. Rev. Lett.}
{\bf 39} 691 (1977).
%
\bibitem{mandelwolfbook} H.J. Kimble and L. Mandel, {\it Phys.
Rev.} A {\bf 13} 2123 1976; see also: L. Mandel and E. Wolf, {\it
Optical Coherence and Quantum Optics} (Cambridge University
Press, Cambridge, 1999).
%
\bibitem{kimble77a}
G.S. Agarwal, {\it Phys. Rev.} A {\bf 37} 1383 (1976); H.J. Kimble
and L. Mandel, {\it ibid.} A {\bf 15} 689 1977; G.S. Agarwal, {\it
ibid.} {\bf 18} 1490 (1978); W. Vogel, D.-G. Welsch and K. W$\rm
\acute{o}$dkiewicz, {\it ibid.} {\bf 28} 1543 (1983).
%
\bibitem{agarwal79} G.S. Agarwal and S.S. Jha, {\it Z. Physik} B
{\bf 35} 391 (1979).
%
\bibitem{braje04} D.A. Braje, V. Bali$\rm \acute{c}$, S. Goda, G.Y. Yin and S.E.
Harris, {\it Phys. Rev. Lett.} {\bf 93} 183601 (2004).
%
\bibitem{harris97} S.E. Harris, {\it Phys. Today} {\bf 50} 36 (1997);
O.A. Kocharovskaya and Y.I. Khanin, {\it JETP Lett.} {\bf 48} 630
(1988).
%
\bibitem{balic05} V. Bali$\rm \acute{c}$, D.A. Braje, P. Kolchin, G.Y. Yin and
S.E. Harris, {\it Phys. Rev. Lett.} {\bf 94} 183601 (2005).
%
\bibitem{kimble03} A. Kuzmich, W.P. Bowen, A.D. Boozer, A. Boca,
C.W. Chou, L.M. Duan, H.J. Kimble, {\it Nature} {\bf 423} 731
(2003).
%
\bibitem{lukin03} C.H. van der Wal, M.D. Eisaman, A. Andre, R.L.
Walsworth, D.F. Phyllips, A.S. Zibrov, M.D. Lukin, {\it Science}
{\bf 301}, 196 (2003).
%
\bibitem{chuangnielsenbook} see e.g., I.L. Chuang and M.A. Nielsen {\it Quantum Computation and Quantum Information}
(Cambridge University Press, Cambridge, 2000).
%
\bibitem{duan01} L.-M. Duan, M.D.
Lukin, J.I. Cirac and P. Zoller, {\it Nature} {\bf 414} 413 (2001)
%
\bibitem{scully05} H. Xiong, M.O. Scully and M.S. Zubairy, {\it Phys. Rev.
Lett.}{\bf 94} 023601 (2005); M.O. Scully, {\it ibid.}{\bf 55}
2802 (1985); M.O. Scully and M.S. Zubairy, {\it Phys. Rev.} A
{\bf 35} 752 (1987), W. Schleich, M.O. Scully and H.-G. von
Garssen, {\it ibid.} {\bf 37} 3010 (1988); W. Schleich and M.O.
Scully, {\it ibid.} {\bf 37} 1261 (1988).
%
\bibitem{scully95} U. Rathe and M. Scully, {\it Lett. Math. Phys.}
{\bf 34} 297 (1995); M.O. Scully, U.W. Rathe, C. Su and G.S.
Agarwal, {\it Opt. Commun.} {\bf 136} 39 (1997).
%
\bibitem{scully04} M.O. Scully and C.H.R. Ooi, {\it Quantum Semiclass. Opt.}
{\bf 6} s816 (2004).
%
\bibitem{agarwal04} G.S. Agarwal, G.O.
Ariunbold, J. von Zanthier and H. Walther, {\it Phys. Rev.} A
{\bf 70} 063816 (2004);  G.S.Agarwal, J. von Zanthier, C. Skornia
and H. Walther, {\it Phys. Rev.} A {\bf 64}, 063801 (2002).
%
\bibitem{rempe05} M. Hennrich, A. Kuhn and G. Rempe, {\it Phys. Rev.
Lett.}{\bf 94} 053604 (2005).
%
\bibitem{carmichael78} H.J. Carmichael, P. Drummond, P. Meystre
and D.F. Walls {\it J. Phys.} A {\bf 11} (1978).
%
\bibitem{harris93} S.E. Harris {\it Phys. Rev. Lett.} {\bf 70}, 552
(1993).
%
\bibitem{agarwal93} G.S.Agarwal, {\it Phys. Rev. Lett.} {\bf 71}, 1351 (1993)
%
\bibitem{beige98} A. Beige and G.C. Hegerfeldt {\it Phys. Rev.} A
{\bf 58} 4133 (1998).
%
\bibitem{mabuchi02} A.J. Berglund, A.C. Doherty and H. Mabuchi, {\it Phys. Rev. Lett.}
{\bf 89} 068101 (2002); G.O. Ariunbold, G.S. Agarwal, Z. Wang,
M.O. Scully and H. Walther, {\it J. Phys. Chem.} A {\bf 108} 2402
(2004).
%
\bibitem{alzar03} C.G. Alzar, L. Cruz, J.A. Gomez, M.F. Santos and
P. Nussenzveig, {\it Europhys. Lett.} {\bf 61}, 485 (2003).
%
\bibitem{motomura05} K. Motomura, M. Tsukamoto, A. Wakiyama, K.
Harada and M. Mitsunaga, {\it Phys. Rev.} A {\bf 71} 043817
(2005).
%
\bibitem{volodya} V.A. Sautenkov, Yu.V. Rostovtsev,
and M.O. Scully, {\it Phys. Rev.} A {\bf 72}, 065801 (2005).
%
\bibitem{yabuzaki91} T. Yabuzaki, T. Mitsui and U. Tanaka, {\it Phys. Rev.
Lett.} {\bf 67} 2453 (1991).
%
\bibitem{walser94} R. Walser and P. Zoller, {\it Phys. Rev.} A {\bf
49} 5067 (1994).
%
\bibitem{mcintyre93} D.H. McIntyre, C.E. Fairchild, J. Cooper and R. Walser, {\it
Opt. Lett.} {\bf 18}, 1816 (1993).
%
\bibitem{camparo98} J.C. Camparo, {\it JOSA} B, {\it 15}, 1177 (1998); J.C.
Camparo and J.G. Coffer, {\it Phys. Rev.} A {\bf 59}, 728 (1999).
%
\bibitem{bahoura01} M. Bahoura and A. Clairon, {\it Opt. Lett.} {\bf 26}, 926,
(2001).
%
\bibitem{martinelli04} M. Martinelli, P. Valente, H.
Failache, D. Felinto, L.S. Cruz, P. Nussenzveig and A. Lezama,
{\it Phys. Rev.} A {\bf 69} 043809 (2004).
%
\bibitem{vassiliev} V.V. Vassiliev, S.A. Zibrov, V.L. Velichansky,
{\it Rev. Sci. Instrum.} {\bf 77}, 013102 (2006)

\bibitem{akulshin90} A.M. Akulshin, V.A. Sautenkov, V.L. Velichansky, A.S. Zibrov and M.V. Zverkov,
{\it Opt. Commun.}{\bf 77} 295 (1990)
%
\bibitem{roos03} P.A. Roos, S.K. Murphy, L.S. Meng, J.L. Carlsten,
T.C. Ralph, A.G. White and J.K. Brasseur, {\it Phys. Rev.} A {\bf
68} 013802 (2003)
%
\bibitem{gardnerzollerbook} C.W. Gardiner and P. Zoller, {\it Quantum
Noise} (Springer-Verlag, Berlin, 2000)
%
\bibitem{ornstein54} G.E. Uhlenbeck and L.S. Ornstein, {\it Selected papers on Noise and Stochastic
Processes}, ed. N. Wax, (Dover Publications, New York, 1954);
N.G. van Kampen, {\it Stochastic Processes in Physics and
chemistry} (North-Holland, Amsterdam, 1981).
%
\bibitem{fox88} R.F. Fox, I.R. Gatland, G. Vemuri, {\it Phys.
Rev.} A {\bf 38} 5938 (1988).
%
\bibitem{wodkiewicz79} K. W$\rm \acute{o}$dkiewicz, {\it J. Math. Phys.} {\bf 20} 45 (1979)
%
\bibitem{fleishhauer94}
M. Fleishhauer and M.O. Scully, {\it Phys. Rev}. A {\bf 49}, 1973
(1994).
%
\bibitem{kitching02} J. Kitching, S. Knappe and L. Hollberg, {\it Appl. Phys. Lett.}
{\bf 81}, 553 (2002); P.D.D. Schwindt, S. Knappe, V. Shah, L.
Hollberg and J. Kitching, {\it Appl. Phys.  Lett.}, 85, 6409,
(2004).
%
\bibitem{budker02} D. Budker, W. Gawlik, D.F. Kimball,
S.M. Rochester, V.V. Yashchuk and A. Weis, {\it Rev. Mod. Phys.}
 {\bf 74} 1153 (2002).
%
\end{thebibliography}
\end{document}